%% file: main-vgtc.tex
\documentclass[electronic]{vgtc}             

\pdfoutput=1\relax                   
\usepackage{graphicx}                
\DeclareGraphicsExtensions{.pdf,.png,.jpg,.jpeg} 

\graphicspath{{figures/}{pictures/}{images/}{./}} 

\usepackage{microtype}                 
\PassOptionsToPackage{warn}{textcomp}  
\usepackage{textcomp}                  
\usepackage{mathptmx}                  
\usepackage{times}                     
\usepackage[superscript,biblabel]{cite}                      

\input{commands}

\usepackage{slashbox}
\usepackage{multirow}
\usepackage{subfig}
\usepackage{array}
\usepackage{tcolorbox}
\usepackage[normalem]{ulem}

\newcolumntype{M}[1]{>{\centering\arraybackslash}m{#1}}
\newcolumntype{L}[1]{>{\raggedright\arraybackslash}m{#1}}
\newcolumntype{N}{@{}m{0pt}@{}}

\onlineid{9717}

\vgtccategory{VAHC}

\vgtcinsertpkg

\title{Looking beyond the horizon: Evaluation of four compact visualization techniques for time series in a spatial context}

\author{Manuel Dahnert\,\thanks{Manuel Dahnert and Johannes Kehrer are with Technische Universit\"at M\"unchen, Germany. E-mail: \{manuel.dahnert, johannes.kehrer\}@tum.de}
\and Alexander Rind\,\thanks{Alexander Rind and Wolfgang Aigner are with St.~Poelten University of Applied Sciences, Austria. E-mail: \{alexander.rind, wolfgang.aigner\}@fhstp.ac.at}
\and Wolfgang Aigner\,\footnotemark[2]
\and Johannes Kehrer\,\footnotemark[1]~\,\thanks{Johannes Kehrer is also with Siemens AG, Corporate Technology, Germany.}}

\abstract{\input{abstract}
} 

\CCScatlist{
  \CCScatTwelve{Human-centered computing}{Visu\-al\-iza\-tion}{Empirical studies in visu\-al\-iza\-tion}{};
  \CCScatTwelve{Human-centered computing}{Visu\-al\-iza\-tion}{Visu\-al\-iza\-tion techniques}{}
}

\begin{document}

\maketitle

\input{introduction}

\input{related-work}

\input{visualization}
\input{user-study}
\input{results}
\input{discussion}

\input{limitation}

\input{conclusion}

\acknowledgments{
This work was supported in part by
the Austrian Science Fund (FWF): P25489-N23 and P27975-NBL via the KAVA-Time and VisOnFire projects, the Austrian Ministry for Transport, Innovation and Technology (BMVIT) under the ICT of the future program via the VALiD project (FFG 845598), as well as the European Union under the ERC Advanced Grant 291372: SaferVis -- Uncertainty Visualization for Reliable Data Discovery.
}

\section*{Declaration of Conflicting Interests}
The authors declare that there is no conflict of interest.

\bibliographystyle{abbrv-doi-hyperref-narrow}
\bibliography{literature}
\end{document}

%% file: commands.tex
\newcommand{\supplURL}{{\url{http://phaidra.fhstp.ac.at/o:3569}}}

\newcommand{\hg}{\textit{HG}}
\newcommand{\bp}{\textit{CBP}}
\newcommand{\chg}{\textit{CHG}}
\newcommand{\bhg}{\textit{BHG}}

\usepackage{todonotes}

\definecolor{improvement}{rgb}{0.8, 1.0, 0.5}
\definecolor{remove}{rgb}{0.92,0.5,0.4}

%% file: abstract.tex
Visualizing time series 
in a dense spatial context such as a geographical map is a challenging task, which requires careful balance between the amount of depicted data and perceptual precision.
Horizon graphs are a well-known technique for compactly representing time series data.
They provide fine details while simultaneously giving an overview of the data where extrema are emphasized.
Horizon graphs compress the vertical resolution of the individual line graphs, but they do not affect the horizontal resolution.
We present two variations of a new visualization technique called \emph{collapsed horizon graphs}
which extend the idea of horizon graphs to two dimensions.
Our main contribution is a quantitative evaluation that experimentally compares four visualization techniques with high visual information resolution (compact boxplots, horizon graphs, collapsed horizon graphs, and braided collapsed horizon graphs).
The experiment investigates the performance of these techniques across tasks
addressing both individual graphs as well as groups of adjacent graphs.
Compact boxplots consistently provide good results for all tasks, horizon graphs excel, for instance, in maximum tasks but underperform in trend detection. Collapsed horizon graphs shine in certain tasks in which an increased horizontal resolution is beneficial. 
Moreover, our results indicate that the visual complexity of the techniques highly affects users' confidence and perceived task difficulty.

%% file: introduction.tex
\begin{figure*}[htbp]
\centering
 	\subfloat[Compact Boxplots\label{subfig:teaser_bp}]
 	{
     \includegraphics[width=0.23\linewidth]{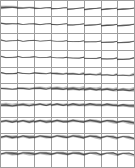} 
 	}   ~
 	\subfloat[Horizon Graphs (HG)\label{subfig:teaser_hg}]
 	{
     \includegraphics[width=0.23\linewidth]{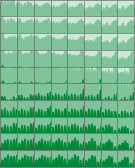} 
 	}   ~
 	\subfloat[Collapsed Horizon Graphs\label{subfig:teaser_chg}]
 	{
     \includegraphics[width=0.23\linewidth]{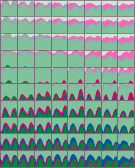} 
 	}   ~
 	\subfloat[Braided Collapsed HG\label{subfig:teaser_bhg}]
 	{
     \includegraphics[width=0.23\linewidth]{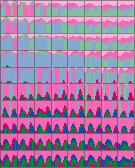} 
 	}
   \caption{Four compact visualization techniques depict the same meteorological data set containing temperature values over time. Each time series represents a geospatial location and consists of 50 time steps. To fit into the limited display space, each graphic is restricted to $18 \times 18$ pixels.
   In our evaluation, we study tasks that address both individual and groups of neighboring graphics.}
   \label{fig:teaser}
\end{figure*}

\section{Introduction}

In many application areas, much of the available data has a relation to time. An efficient and effective visualization of time series data has to consider the capabilities of the human visual system as well as the specific properties of the temporal dimension such as linear or cyclic progression\cite{aigner08}. In addition, time series data is often related to space resulting in spatio-temporal data such as weather forecasts, employment rates, or real estate prices.
One important characteristic of such data is that ``everything is related to everything else, but near things are more related than distant things.''\cite{tobler70firstLaw}
Identifying these complex relations both in space and time is an important analytic task.
As data is getting more complex and grows in size and spatial density, there is need for efficient visualization techniques that allow users, for example, to identify spatio-temporal patterns such as extreme values or trends; to compare spatially neighboring time series and identify similarities and differences; and to accurately read the time-dependent values at different time steps (and locations) from the visualization.

Line plots are one of the most common techniques for visualizing time series
but they are not well suited to convey the spatial relationships of the data. 
Geospatial data, on the other hand, is often shown with cartographic maps, following a set of well-established guidelines.\cite{Dykes05Geovisualization}
In order to analyze both the spatial and temporal characteristics of the data,
it has been shown beneficial to directly integrate both aspects in a single visualization, for example, by placing miniaturized versions of line plots onto a map.\cite{andrienko04, fuchs04}
In such a scenario, however, the available display space quickly becomes a limiting factor,
especially when longer time series need to fit into a dense spatial context.

Sparklines,\cite{tufte_2006_beautiful} for example, can be be shrunken down to the height of a text character and still convey information about temporal patterns.
However, the loss of resolution impairs the user's ability to precisely read the values or slopes at specific time steps from the visualization.
Other strategies to satisfy the display requirements in a dense spatial setting are \emph{visual compression} and \emph{data aggregation}.
Visual compression aims at representing the same amount of data in less display space by using additional visual channels or special encodings.\cite{saito05, reijner08, javed10}
Data aggregation, on the other hand, typically computes summary statistics for specific time spans or locations and visualizes these measures instead of the original data.\cite{bade04}

With this article we provide the following contributions:
\begin{itemize}
	\item As main contribution, we present the results of a quantitative user study that compares four compact visualization techniques for time series in a dense spatial context (see \autoref{fig:teaser}): compact boxplots,\cite{bade04} horizontally-downscaled horizon graphs,\cite{saito05, reijner08} as well as two new techniques called collapsed horizon graphs and braided collapsed horizon graphs. 
Our empirical study examines differences in user performance between techniques using data aggregation, color encoding, and size shrinking across a broad range of tasks.

	\item  As secondary contribution, we present collapsed horizon graphs, a technique that extends the idea of horizon graphs to two dimensions. It preserves the horizontal resolution of the data and emphasizes the progression of the time series, such as trends.
We also apply the idea of braided graphs\cite{javed10} in a variation of our technique to avoid occlusions when overlaying graphical elements.

	\item Finally, we present a systematic set of evaluation tasks for user studies that address the special properties of both time and space based on conceptual frameworks of Andrienko and Andrienko \cite{andrienko06book} and Peuquet \cite{peuquet94triad}.
    These tasks focus on spatio-temporal trends and patterns in groups of adjacent graphics.

\end{itemize}

In the following, we will revisit related user studies on time series as well as dense visualizations.
The section \nameref{sec:compactVis} presents the techniques we were evaluating.
In \nameref{sec:userStudy} 
the hypotheses, tasks and the general setup of the user study are described.
Sections \nameref{sec:results} 
and \nameref{sec:discussion} 
present and discuss the results of the user study.
Section \nameref{sec:limitations} 
points to limitations of the evaluation and possible directions for future work. The last section concludes the article.

%% file: related-work.tex
\section{Related work}\label{sec:relatedWork}

The design space for visualizations is huge and it can be hard to find an effective and efficient design.\cite{munzner_2014_book}
For example, Aigner et al.\cite{aigner11} survey over 100 visualization techniques for time-oriented data, of which many additionally support spatial data.
While some design choices can be based on guidelines or the existing body of research on human visual capabilities, for other choices one needs to consider trade-offs in respect of the addressed user need.
In such situations, empirical data from user studies can yield decisive insight.

A number of user studies has investigated the performance of dense visualization techniques, visualization techniques for time series, or combinations thereof.
Liu and Shen\cite{liu15} have used a large and dense setting of numerous dynamic networks to evaluate different matrix visualization techniques in two controlled experiments.
Albers et al.\cite{albers14} have compared position-based and color-based visualization techniques of time series by conducting a crowd-sourced evaluation. 
Line plots with indexing have been compared to other techniques for visualization of time series with heterogeneous value ranges.\cite{aigner_2011_bertin}
Aigner et al.\cite{aigner_2012_comparative} have compared two techniques that supplement time series line plots with qualitative abstractions of the time series.
Several studies\cite{albo_2016_static, griffin_comparison_2006, robertson_2008_effectiveness} have compared variants of small multiples with animation.
Bauer et al.\cite{bauer_2010_design} compared compact sparklines to a larger table displays in a medical setting.
Lam et al.\cite{lam07} have investigated different conditions of overview use and multiple visual information resolutions.

Analyzing temporal glyphs in a small multiple setting is a common task across various areas of application. Fuchs et al.\cite{fuchs13} have investigated the performance and accuracy of four different visualization techniques in a controlled experiment. 
The experiment by Javed et al.\cite{javed10} involved 
line plots, juxtaposed line plots, horizon graphs, and braided graphs. 
In another study, horizon graphs have been compared to qualizon graphs,\cite{federico14} which have supplemental qualitative abstractions.
Heer et al.\cite{heer09}\ have tested different configurations of horizon graphs to identify which vertical resolution is still effective.
The study by Perin et al.\cite{perin_2013_interactive} experimentally compared horizon graphs with and without interaction as well as compact line plots.

However, none of these user studies has involved dense visualization in respect of the horizontal resolution. Neither have there been study designs focusing on spatial neighborhood of small multiples.

%% file: visualization.tex
\section{Compact time series visualization}
\label{sec:compactVis}
In this section, we recall horizon graphs and compact boxplots, two existing compact visualization techniques that use visual compression and data aggregation to cope with the high density of the data. We then present and discuss the design rationals of our extension to horizon graphs.  \autoref{tab:properties} summarizes the properties of the techniques.

\input{boxplot}
\input{horizongraph}

\input{colhorgraph}

\input{braidedgraph}

\subsection{Visualization properties}
\begin{table}[hbt]
	\small\sffamily\centering
	\begin{tabular}{|L{0.7cm}|L{0.6cm}|L{2.0cm}|L{1.3cm}|L{1.7cm}|N}
	\hline
	\rule{0pt}{2ex}\textbf{Tech.}&\rule{0pt}{2ex}\textbf{Axis} &\rule{0pt}{2ex}\textbf{Type} &\rule{0pt}{2ex}\textbf{Slopes} &\rule{0pt}{3ex}\textbf{Visual clutter}\rule{0pt}{3ex}\\[3pt]
	\hline
	\multirow{2}{*}{\bp} & \multirow{2}{*}{H} & V: None & \multirow{2}{*}{Flat} & \multirow{2}{*}{Low} \\
	& & H: Aggregation & & \\
	\hline
	\multirow{2}{*}{\hg} & \multirow{2}{*}{V} & V: Color & \multirow{2}{*}{Steep} & \multirow{2}{*}{Medium} \\
	& & H: None & & & \\
	\hline
    \multirow{2}{*}{\chg} & \multirow{2}{*}{V, H} & V: Color & \multirow{2}{*}{Preserved} & \multirow{2}{*}{High} \\
	& & H: Color & & \\
	\hline
	\multirow{2}{*}{\bhg} & \multirow{2}{*}{V, H} & V: Color & \multirow{2}{*}{Preserved} & \multirow{2}{*}{Very High} \\
	& & H: Color & & \\
	\hline
	\end{tabular}
	\caption{Classification of the different visualization techniques.\\
    V: Vertical, H: Horizontal.}
	\label{tab:properties}
\end{table}

\autoref{tab:properties} summarizes the visual properties of the techniques used in our study.  These properties are briefly described in the following:

\begin{itemize}
	\item \textbf{Compression axes:} \quad In relation to the type of compression, this describes the axes that are used to compress the data. 
	\item \textbf{Types of compression:} \quad In this user study we consider visual compression, i.e. color encoding and a subsequent collapsing, data aggregation and horizontally shrinking of the visual representation.
	\item \textbf{Graph slopes:} \quad Depending on the type of compression, the slopes of the time series in the compact representation may change compared to the original line graph. Since the perception of graph slopes is an important part of a visual analysis, this factor takes the steepness or flatness of the slope into account.
	\item \textbf{Degree of visual clutter:} \quad Depending on the number of used colors and the frequency of color changes, the visual outcome may become highly cluttered and therefore difficult to analyze.
\end{itemize}

So far no empirical evidence is available from prior studies that would compare these techniques across a range of different user tasks and provide a better understanding of which techniques are better suited for which kind of task and what the role of the different properties is in this regard. To close this gap, we performed a comparative empirical study and analyzed the gathered results. Next, we will describe our study design, the used tasks as well as the study procedure.

%% file: boxplot.tex
\subsection{Compact Boxplots (CBP)}

\begin{figure}[htbp]
	\centering
	\includegraphics[width=0.7\columnwidth]{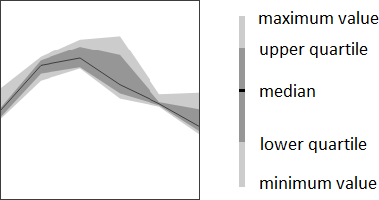}
	\caption{Compact boxplot visualization showing computed summary statistics over time (compare to Bade et al.~\cite{bade04}).}
	\label{fig:bp_1}
\end{figure}

Boxplots are widely used in many scientific disciplines to represent distributions of data values.
Bade et al.~\cite{bade04} suggested a compact representation of time-dependent data based on a redesign of boxplots by Tufte~\cite{tufte}.
This \emph{compact boxplot} (\bp{}) resembles a traditional line graph surrounded by quartile bands. 
It splits the line graph into non-over\-lapping time intervals of equal duration and computes summary statistics---the median, upper and lower quartile, as well as minimum and maximum value per time interval. The statistics are then depicted instead of the original line graph as shown in Fig.~\ref{fig:bp_1}.
While visual clutter is quite reduced by this technique,
even when many time steps fall into an aggregation interval, 
high-frequency details can be lost due to the data aggregation. Note that a compact boxplot resembles a line graph if the variance of the data is low.

%% file: horizongraph.tex
\subsection{Horizon Graphs (HG)}

\begin{figure}
	\centering
    \subfloat[\label{subfig:hg_1}]
	{\includegraphics[width=0.4\linewidth]{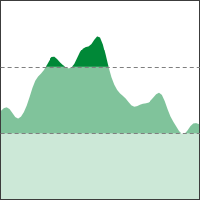}}
      \quad \quad \hphantom{\includegraphics[width=0.13\linewidth]{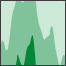}} \\
    \subfloat[\label{subfig:hg_2}]
	{\includegraphics[width=0.4\linewidth]{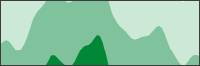}} \quad \quad
    \subfloat[\label{subfig:hg_3}]
   	{\includegraphics[width=0.13\linewidth]{hg_3}} 
    \caption{A line graph is divided into three horizontal bands and colored with a sequential color scheme~(a). The bands are then overlaid vertically~(b). To fit the spatial requirements the graph is shrunk horizontally~(c). We omit the idea of mirroring or offsetting negative values at the zero line as described in \nameref{sec:userStudy}.
    } 
    \label{fig:hg}
\end{figure}

Reijner~\cite{reijner08} introduced \emph{horizon graphs} (\hg), which adapt the concept of two-tone pseudo coloring by Saito et al.\cite{saito05}
\hg{} reduce the vertical display space of a line graph without losing precision when reading data values at a specific time step, thus increasing the overall data density.
They divide a given line graph into non-overlapping horizontal bands of equal height (see \autoref{subfig:hg_1}).
Starting from the zero line, the bands are then colored with a diverging color scheme, which uses different hues for positive and negative values.
Negative values are then typically mirrored at the zero line and superimposed with the positive values. 
Finally, the bands are overlaid from top to bottom, such that bands with higher values are shown in front of those with lower values (see \autoref{subfig:hg_2}).
By overlaying the horizontal bands, the slopes of the original line graph are maintained.
This compact visualization technique also emphasizes maximum and minimum values, since these pop-out due to the color coding and overlay.
Heer et al.~\cite{heer09} investigated the difference between offsetting negative values instead of mirroring them in a user study. By offsetting the values beneath the zero line, the intuitive decline of negative values is retained. However, they found no significant difference in accuracy and completion time.

For our study, we omit the idea of mirroring and offsetting negative values, because of the non-zero value domain of the time-dependent axis, see Section \nameref{sec:design_choices}. Further, we horizontally shrink the horizon graph as shown in \autoref{subfig:hg_3} to fit also longer time series into a dense spatial context. The slopes of the original line graph thereby get steeper, which may affect the judgment of temporal trends \cite{Talbot11}.

%% file: colhorgraph.tex
\subsection{Collapsed Horizon Graphs (CHG)}

\begin{figure}[htbp]
	\centering
    \subfloat[\label{subfig:chg_1}]
	{\includegraphics[width=0.4\linewidth]{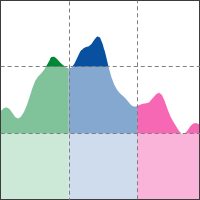}}
    \quad \quad \hphantom{\includegraphics[width=0.13\linewidth]{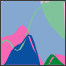}} \\
    \subfloat[\label{subfig:chg_2}]
	{\includegraphics[width=0.4\linewidth]{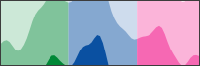}} \quad \quad
    \subfloat[\label{subfig:chg_3}]
   	{\includegraphics[width=0.13\linewidth]{chg_3}} 
	\caption{A line graph is divided into three horizontal bands and three vertical slices, respectively, and colored with a bivariate color map~(a). The bands are first overlaid vertically~(b) and then collapsed horizontally from left to right~(c). 
    Occluded parts are shown as contours.}
	\label{fig:chg}
\end{figure}

\textit{Collapsed horizon graphs} (\chg) adapt the concept of horizon graphs by extending the idea of overlaying discrete intervals to both dimensions. The original line graph is first divided into $B$ horizontal \emph{bands} of equal height and $S$ vertical \emph{slices} of equal width, which yields $B \times S$ cells. The cells are then colored using a bivariate color map \cite{brewer94,Bernard15} which encodes two variables simultaneously as explained below.
Similar to \hg, the colored cells are then overlaid from top to bottom. Again, cells with higher values occlude cells with lower values as shown in \autoref{subfig:chg_2}. To reduce display space, the cells are finally collapsed horizontally in a specific order per band, for instance, from left to right.
In order to preserve the shape information from cells that may become occluded, the occluded parts thereby ``shine through'' as contour lines in the color of the corresponding cell (see \autoref{subfig:chg_3}).
The result is a compact representation of the original line graph, which retains most of its original visual information resolution \cite{lam07}. {\chg} only requires $^1/_B$ of the height and $^1/_S$ of the width of the line graph, where it is still possible to read out the data value at a specific time step with high precision.
The selection of an appropriate bivariate color map is crucial for our technique and will be discussed in the following.

\begin{figure}[htbp]
	\centering
	\includegraphics[width=\columnwidth]{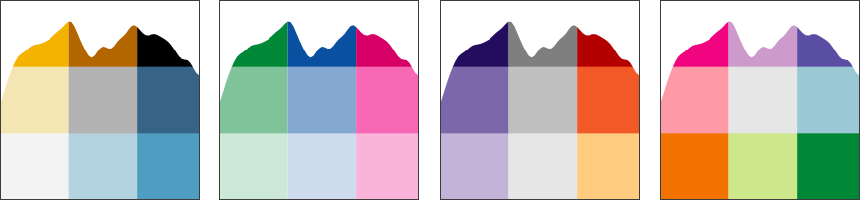}
	\caption{We have examined different bivariate color maps while designing \chg{} and \bhg. From left to right: Sequential--sequential, sequential--qualitative, sequential--diverging, diverging--diverging.\cite{brewer94}}
	\label{fig:chg_colormap}
\end{figure}

\paragraph{Color selection.}
In contrast to horizon graphs, we omit the idea of mirroring or offsetting negative values to reduce both the visual complexity and the number of colors. Consequently, we use a sequential color scheme for encoding the vertical position in the original line graph as shown in \autoref{subfig:chg_1}. By introducing also a horizontal collapsing step in \chg, we need to extend this color scheme to a bivariate color map \cite{brewer94, Bernard15} in order to assign each cell a unique color.
We initially tested different variations of bivariate color maps (see \autoref{fig:chg_colormap}).
When using a sequential color scheme for the horizontal position in the line graph (\autoref{fig:chg_colormap}a), later time steps are visually emphasized. 
In contrast, a diverging color scheme does not fit the characteristics of a linear time axis which usually has no center (\autoref{fig:chg_colormap}c and \ref{fig:chg_colormap}d).
Therefore, we chose a qualitative color scheme for the horizontal position in the original line graph, because it divides the time axis into equally perceivable intervals (\autoref{fig:chg_colormap}b).

\begin{figure}[htbp]
	\centering
	\subfloat[\label{subfig:lr}]
	{\includegraphics[width=.35\columnwidth]{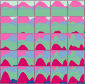} }   \quad
	\subfloat[\label{subfig:rl}]
	{\includegraphics[width=.35\columnwidth]{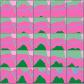} }
	\caption{Comparison between different orderings of slices for \chg. With a right to left order increasing slopes are highlighted~(a), whereas decreasing slopes are emphasized with  a left to right order~(b).} 
	\label{fig:rl_lr}
\end{figure}

\paragraph{Ordering of slices.} \quad
One important property of \chg{} is the highlighting of a particular time-series interval. In the collapsing step, each horizontal band is collapsed individually with a particular order. Since cells in the foreground may occlude other cells that are only represented by a contour, they are more dominant in the visualization.
Depending on the analysis task at hand, therefore, either the first or last time interval can be emphasized by choosing an appropriate order.
Consequently, either increasing or decreasing temporal trends can be emphasized by showing the corresponding layer with solid color (see \autoref{fig:rl_lr}).

\paragraph{Overview \& Detail.} \quad
\chg{} provide both overview and detail over the given data, which facilitates the analysis of large spatio-temporal data grids.
With the help of highlighting a particular time interval through selecting the appropriate ordering of the slices certain properties of the dataset can be pre-attentively grasped. 

The first property is the temporal trend, as described above. If for instance a certain area of the dataset appear redish (cf. right-most column in bivariate color map \autoref{fig:chg_colormap}b) then this indicates an increasing trend in this area.

Similar to \hg{} if a certain area of the data grid appears e.g. with more saturated colors (cf. \autoref{fig:chg_colormap}b) then this indicates a generally higher value in this region.

If an individual time series is to be examined in greater detail, then the \chg{} can be mentally unfolded again. Previously occluded parts can be reconstructed with the help of the contour lines.

%% file: braidedgraph.tex
\subsection{Braided Collapsed Horizon Graphs (BHG)}

\begin{figure}[htbp]
	\centering
    \subfloat[\label{subfig:bhg_2_1}]
	{\includegraphics[width=0.2\linewidth]{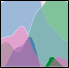}}
      \quad \quad
    \subfloat[\label{subfig:bhg_2_2}]
	{\includegraphics[width=0.2\linewidth]{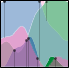}}
	\quad \quad
    \subfloat[\label{subfig:bhg_2_3}]
	{\includegraphics[width=0.2\linewidth]{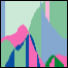}}

	\caption{The cells of \chg{} are layered and intersection points are identified. At these intersection points the segments are then ordered such that visibility is ensured.}
	\label{fig:bhg_2}
\end{figure}

While collapsed horizon graphs use contour lines to convey the information of occluded parts, the concept of braided graphs can also be applied.\cite{javed10}
Braided graphs overlay multiple colored line graphs on top of each other (\autoref{subfig:bhg_2_1}) and search for intersection points, i.e., points where the data value of different line graphs cross (\autoref{subfig:bhg_2_2}). At each intersection point, the filled areas below the line graph are then split into segments. Segments with a higher value are then drawn behind segments with a lower value (\autoref{subfig:bhg_2_3}). This ensures that each line graph segment is visible at every point in time.

Applying the idea of braiding the individual cells of a collapsed horizon graph ensures the visibility of every point in time across the whole data set.
A \textit{braided collapsed horizon graph} (\bhg) is obtained by first collapsing the 2D colored line graph in both dimensions. Second, the cells are braided as described above. This removes the necessity of collapsing the slices in a specific order.
Hence, the chosen order does not imply a focus on increasing or decreasing trends and the whole line graph is perceived equally.
However, depending on the number of intersections in the depicted data, {\bhg} can suffer from a high amount of visual clutter.

%% file: user-study.tex
\section{User study}\label{sec:userStudy}

The goal of this user study is to compare the effects of data aggregation and visual compression techniques for time series in a dense spatial context. Since some techniques incorporate high visual complexity and cognitive load, we examine these attributes and investigate if users can accurately read data properties from the visual encodings in a reasonable amount of time.

\subsection{Hypotheses}
Based on the specifics of the presented techniques as well as prior research, we formulate five hypotheses that guide our research. 

\begin{itemize}
\item[H1:] \textit{Participants utilizing techniques with only one compression axis (\hg{} and \bp) are generally faster than using a technique with two compression axes (\chg{} and \bhg).}
We assume that mentally unfolding an additional axis requires a significant amount of time which leads to longer completion times.
\item[H2:]
\textit{Participants using techniques with two compression axes perform better at synoptic tasks, i.e., tasks involving sets of data elements.}
Since different time intervals are encoded in color in {\chg} and {\bhg}, we assume that it should be easier for the user to identify similar temporal patterns in spatially neighboring time series. Accordingly, characteristics such as maximum values or similar slopes should be easier to identify,  
which makes it faster to get an overview of the data.
\item[H3:]
\textit{Participants using visual compression techniques (\hg, \chg{} and \bhg) can read out values more accurately than those using aggregation techniques (\bp).}
Visual compression will preserve high-frequency details of the time series, which will lead to more accurate results.
Aggregation will lose these details and hence will lead to less accurate results.
\item[H4:]
\textit{Participants using techniques with two compression axes (\chg{} and \bhg) are more accurate
but slower at stating the temporal position of a certain event than using \hg.}
Additionally, we assume that these techniques are more accurate when searching for the temporal position of a given value.
\item[H5:]
\textit{Participants using techniques with two compression axes will report better
performance at slope tasks.}
Collapsing the time series in the horizontal dimension will preserve the slope of the time series. 
Hence estimating the slope will not be negatively affected and large changes will pop-out.
\end{itemize}

\input{tasks}

\subsection{Experimental conditions}\label{sec:conditions}

In the following, we describe the conditions of our experiment.
\begin{itemize}
	\item \textbf{Visualization technique (V):} \quad Horizon Graphs (\hg), Compact Boxplots (\bp), Collapsed Horizon Graphs (\chg) and Braided Collapsed Horizon Graphs (\bhg).
	\item \textbf{Task (T):} \quad All tasks, except Task 03, are performed with all techniques. Task 03 is only performed with \chg, since the ordering of the slices may emphasize positive or negative slopes.
	\item \textbf{Repetitions (R):} \quad All tasks, except Task 08, are repeated 2 times with different datasets to increase robustness. Task 08 is repeated 3 times, where one repetition queries the whole time interval (0:00-24:00) of the 24-hours metaphor, see Section~\nameref{sec:design_choices}, one repetition tests an interval used by {\chg} and {\bhg} (e.g., 16:00-24:00), and one tests an arbitrary 8h time interval (e.g., 12:00-20:00).
\end{itemize}
The design of the evaluation is a full factorial within-subject design with $\left|V\right| \times \left|T\right| \times R$ factors, which yields $4 \times 8 \times 2 + 4 \times 1 \times 3 + 1 \times 1 \times 2 = 78$~conditions.
The order of the techniques is random for each participant.
For each technique, the tasks are then presented in the order as described in Section~\nameref{sec:tasks} to the participant.

\subsection{Design choices}\label{sec:design_choices}

\begin{itemize}
	\item \textbf{Synthetic data generation:} \quad 
    Similar to other user studies,\cite{javed10,lam07,fuchs13, heer09} we use synthetic data instead of real-world data to have better control of the visual outcome and data features and to ensure equal task difficulty across the data sets.
    First, we generate a set of time series, each consisting of 72 data points, using a random walker algorithm and a subsequent smoothing (compare to Heer et al.~\cite{heer09}).
    To consider also the relationships of spatially neighboring time series, each time series represents a weighted sum between the previously generated time series and a newly generated one (influence of the previous time series is 25\%). For each task and technique, we then generate the required number of time series, which are spatially arranged in a grid using a space-filling Hilbert curve.
    Our algorithm also assures task-specific requirements, e.g., the data set of Task 2 contains at least one time series with an increasing slope.
These requirements to the data generation should ensure similar difficulty of the data sets but also create realistic time series.
For each repetition, we generated 3 different datasets from which one dataset was drawn by random for each participant.

    \item \textbf{Data domain:} \quad In our user study, the time steps follow a 24-hours metaphor and the time-dependent values range from 0--100. Both value ranges and units are common in spatio-temporal scenarios, e.g., percentage rates and 1-day forecasts.
    
    \item \textbf{Omit mirroring of negative values:} Restricting the data values to the positive domain removes the necessity of having a zero line and thus reduces the number of required colors in case of \hg, \chg{} and \bhg. Addressing negatives values was beyond the scope for this user study.

    \item \textbf{Graph size:} \quad Each time series uses only $24\times 24$ pixels to fit the spatial requirements. This is based on earlier results from Heer et al.\cite{heer09} and Javed et al.\cite{javed10}
    \item \textbf{Compression rates:} \quad All visual compression techniques use three bands to compress the vertical resolution. \chg{} and \bhg{} additionally divide the horizontal axis in three slices. \bp{} aggregates over three time steps.
    \item \textbf{Static representation:} \quad Similar to Javed et al.~\cite{javed10} our evaluation only considers static representation of the techniques and does not offer any interaction with the displayed graphs, such as animation or tool tips.
\end{itemize}

\subsection{Pilot study}
We performed a pilot study with two participants. The participants were told to complete the study based on the written instructions. They were encouraged to speak their thoughts and comment if the tasks or the instructions were unclear (think aloud protocol). Based on their feedback, we lowered the number of repetitions per task from three to two to avoid fatigue of the participants and to reduce the overall time to complete the study to less than an hour. Additionally, we changed the shape of the time markers from a line to an up-pointing triangle to increase their visibility and to facilitate the perception of the color of the indicated time interval.

\subsection{Participants}
We recruited 124 unpaid participants to take part in our user study.
After validating the submitted data, i.e. ensuring a reasonable participation time and effort, e.g. minimum of 30 minutes of participation time), and verifying its completeness, results of 100 participants remained.
From those 100 people, 55 reported to be male, 44 female and 1 participant did not specify it. The average age was 21 years with a minimum of 18 years and a maximum of 37 years. 62 people reported to have normal vision and 38 have corrected to normal vision. 
Most of the participants were students from a lecture in which the participation in the experiment was offered as an alternative to another course assignment. Performing the assignment and taking part in the user study required a similar amount of time and effort.

\subsection{Procedure}

The evaluation prototype is based on the EvalBench \cite{aigner13} framework and was adapted to fit the needs of this user study. 
The study material was provided to the study participants as download on a webpage to allow for a large number of subjects. %
After downloading the packaged Java application and finishing all tasks, a result log file was generated, which was then anonymously uploaded to a file server.
While we were able to reach a large number of subjects, equal conditions cannot be guaranteed during the user study as participants performed the experiments on their own devices in their own setup. However, we share this circumstance with the increasingly popular use of Amazon's Mechanical Turk for conducting user studies. 
The first part of the user study included a self-reporting section with demographical questions (age, gender, educational degree, familiarity with time series) as well as questions concerning the visual capabilities, e.g. color blindness, vision correction, etc.

The actual controlled experiments consisted of multiple parts. 
In the first part, the current visualization technique was presented and the user could solve training tasks to get familiar with the technique as well as with the evaluation software itself. After each training task, the user saw the correct result and could decide whether to continue with training, see the explanation of the technique again, or start the actual evaluation.
In the second part, the user had to complete the presented tasks as accurately but also as fast as possible. Each task consisted of multiple repetitions, where the user could also decide to skip a repetition. To prevent learning effects during the same task and across techniques, we use different data sets for each repetition. If the task utilized a marker to indicate a certain point in time, then the position of the marker changes with each repetition.
After each task the user had to state his/her confidence in the given answers and rate the ease-of-use of the technique for the particular task
on a 7-point Likert scale.

Each participant solved all tasks for a particular technique and then proceeded to the next technique.
While the order of the tasks was fixed and presented for each technique as described in Section~\nameref{sec:tasks}, the order of the individual techniques were counter-balanced using a Latin square ordering scheme. 
All necessary data, such as exemplary pictures of the evaluation software, collected results, and analysis scripts can be found as supplementary material at \supplURL{} to ensure reproducibility of the user study and its results.

%% file: tasks.tex
\subsection{Tasks}\label{sec:tasks}

To systematically evaluate our hypotheses and to
cover the specific properties 
of spatio-temporal data, we derive a set of tasks using the conceptual frameworks by Andrienko and Andrienko \cite{andrienko06book} and Peuquet.\cite{peuquet94triad} 
The former distinguish between two categories of tasks: \emph{Elementary tasks} focus on individual elements of the data.
\emph{Synoptic tasks}, on the other hand, operate on a broader extent (temporal as well as spatial). 

For instance, looking up the value at a specific time point within a time series or comparing the relationship between two time points are considered elementary. Searching for patterns within sets of time series or analyzing all time points of a specific time series are examples of synoptic tasks.

The synoptic level unifies the \textit{intermediate} and \textit{overall} reading levels of Bertin's categorization \cite{bertin83} and the spatio-temporal extension of Koussoulakou and Kraak.\cite{koussoulakou92} In the temporal dimension, the intermediate level asks for changes within a certain time interval, for example. Spatially, this level considers neighboring time series as a whole rather than individual time series. The overall reading level for example takes the whole time series into account or considers all time series of the data set.

The Triad framework by Peuquet \cite{peuquet94triad} utilizes different combinations of ``what'', ``when'' and ``where'' to formulate a task with respect to spatio-temporal data. These questions refer to the objective, as well as the temporal and spatial aspects of a task, respectively.

In the following, we describe the properties and objectives of each of our tasks in more detail.
While Tasks 01--06 have been used in earlier studies~\cite{fuchs13, heer09, javed10} and help to achieve comparability with these studies,
Tasks 07--10 are novel and examine further properties of spatio-temporal data.
Instead of comparing properties of individual time series such as the slope, the user has to consider these properties for groups of neighboring time series (see \autoref{fig:synoptic_slope_task}).

For all tasks in our study, except for Task 04 and 05, multiple time series are depicted in a small grid, which represents the spatial locality of the time series. The grid can also facilitate the analysis of a single time series by considering its surrounding time series.

The categorization of our tasks in both frameworks \cite{andrienko06book, peuquet94triad} is shown in \autoref{tab:task-cat1} and \autoref{tab:task-cat2}.

\begin{table}
	\small\sffamily\centering
	\setlength\tabcolsep{4pt}
    \begin{tabular}{|L{0.5cm}|M{1.8cm}|M{1.8cm}|M{1.8cm}|M{1.21cm}|N}
		\hline
		 \multicolumn{2}{|c|}{\multirow{2}{*}{\backslashbox{\textbf{Time}}{\textbf{Space}}}} 
		& \multirow{2}{*}{\textbf{Elementary}} & \multicolumn{2}{c|}{\textbf{Synoptic}} & \rule{0pt}{3ex} \\[5pt]	
		\multicolumn{2}{|c|}{}	&		& \textit{Intermediate}		& \textit{Overall} & \\[2pt]
		\cline{1-5}
		 \multicolumn{2}{|l|}{\textbf{Elementary}}	&	T04, T05 	&	 			& T01 &\\[10pt]		
		\cline{1-2}\rule{0pt}{2ex} \parbox[t]{0mm}{\multirow{2}{*}{\rotatebox[origin=c]{90}{\textbf{Synoptic}}}} 	
		&	 \multicolumn{1}{l|}{\textit{Intermediate}}										 &	 				& T08		& T07 &\\[15pt]
		& \multicolumn{1}{l|}{\textit{Overall}}											 & T06, T09 	& T10		& T02, T03 &\\[15pt]
		\hline
	\end{tabular}
	\caption{Categorization of the tasks using the elementary-synoptic framework\cite{andrienko06book} and using the spatio-temporal reading levels. \cite{koussoulakou92} The chosen set of tasks covers most of the possible combinations.}
	\label{tab:task-cat1}
\end{table}

\begin{table}%
	\small\sffamily\centering
	\begin{tabular}{*{3}{|l}}
		\hline
		\textbf{Query} & \textbf{Tasks} \rule{0pt}{3ex}\\[3pt]
		\hline
		What + When $\rightarrow$ Where & T01, T02, T03, T07, T08, T10 \rule{0pt}{3ex}\\[3pt]
		What + Where $\rightarrow$ When & T06 \rule{0pt}{3ex} \\[3pt]
		When + Where $\rightarrow$ What & T04, T05, T09 \rule{0pt}{3ex} \\[3pt]
		\hline
	\end{tabular}
	\caption{Task categorization according to the Triad framework.\cite{peuquet94triad}}
	\label{tab:task-cat2}
\end{table}

\begin{table*}[t]
	\small\sffamily\centering
	\begin{tabular}{*{15}{| l}}
	\hline
	\textbf{Task} 						& \textbf{Data size} 		& \textbf{Temporal focus} 					& \textbf{Type of answer} \rule{0pt}{3.5mm}\\[1pt]
	\hline
	T01: Maximum 					& 3 $\times$ 3 graphs 				& Local $^1$							& Single graph \rule{0pt}{4mm}\\
	T02: Slope increasing 			& 3 $\times$ 3 graphs 				& Global	 						& Single graph \\
	T03: Slope decreasing			& 3 $\times$ 3 graphs 				& Global 	  						& Single graph \\
	T04: Discrimination				& 1 $\times$ 2 graphs 				& Local $^2$						 	& Single graph \\
	T05: Difference estimation 		& 1 $\times$ 2 graphs 				& Local $^2$						 	& Value input \\
	T06: Time estimation 			& 3 $\times$ 3 graphs, 1 highlighted & Global 	 						& Time slider \\
	T07: Synoptic search 			& 5 $\times$ 5 graphs 				& Range 								& Multiple graphs \\
	T08: Synoptic slope 			& 9 $\times$ 9 graphs, 3 $\times$ 3 quadrants	& Range 	  						& Single quadrant \\
	T09: Classification 			& 5 $\times$ 5 graphs, 1 highlighted	& Range 	  						& Yes / No answer \\
	T10: Homogeneity 				& 9 $\times$ 9 graphs, 3 $\times$ 3 quadrants	& Global 	  						& Single quadrant \\[1.5pt]
	\hline
	\end{tabular}
	\caption{Overview of the set of tasks. The tasks depict different number of time series to test the performance of the techniques in different configurations: 1 $\times$ 2 for direct comparison, 3 $\times$ 3 to test the overall spatial level and 3 $\times$ 3 quadrants to consider the intermediate level in the spatial dimension.  In Task 6 and Task 9 one graph is highlighted to define the spatial position (``where'') within the data set.\\ $^1$ Markers at same position, $^2$ Markers at different positions.}
	\label{tab:tasks}
\end{table*}

\subsubsection*{Task 01 -- Maximum:}
\begin{figure}[hbt]
	\centering
	\includegraphics[width=0.7\columnwidth]{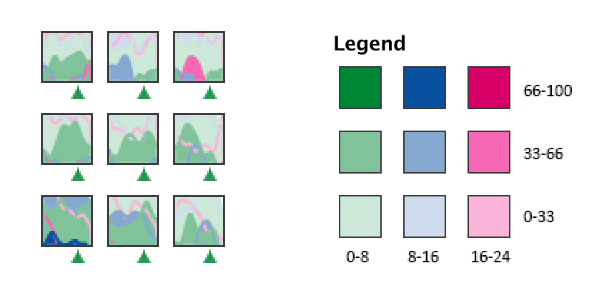}
    \caption{Task 01 with \chg, on the left the 3 $\times$ 3 grid of graphs and time markers beneath are shown. The position of the markers vary each trial. In case of \chg{} and \bhg{} the color of the marker indicates the specific interval. The corresponding legend is depicted on the right.}
    \label{fig:task_01}
\end{figure}

In this task the user is shown a set of 3 $\times$ 3 graphs arranged in a grid and has to detect the time series with the highest value at a specific point in time. The time step in question is indicated by a small marker beneath each time series.
In case of \chg{} and \bhg{}, the marker is also colored in the corresponding hue of the slice to which the point in time belongs.
The position of the markers changes with each repetition (see \nameref{sec:conditions}), but is identical for all graphs within the same repetition, (see \autoref{fig:task_01}).
This task requires to compare the value of all nine graphs at the specified point in time, which is an essential task in time series analysis. 

\subsubsection*{Task 02 -- Increasing slope:}
In this task the time series with the highest increase over the entire timespan has to be found.
Accordingly, the user has to compare the slopes of the time series in the 3 $\times$ 3 grid, in which at least one time series has an increasing slope.
Similar to Javed et al.,\cite{javed10} we approximate the slope by computing the difference between the first and last time step of a time series. While this measure does not consider peaks in the middle of the time series, it is more intuitive and easier to present than comparing the slopes of estimated regression lines, for example.
This task also investigates the relationship between the aspect ratio of a depicted time series and its slope perception.\cite{cleveland88}

\subsubsection*{Task 03 -- Decreasing slope}
With \chg{}, different parts of the time series are emphasized, depending on the order in which the slices are layered.
In case of a left-to-right order, the leftmost slice is in front of the others. Increasing slopes are thus highlighted, while decreasing slopes are only visualized with a contour line. We hypothesize that \chg{} may perform differently for increasing and decreasing slopes, while the other techniques are not affected.
Therefore, we run this task only for \chg{} where we ensure that at least one time series has a decreasing slope.

\subsubsection*{Task 04 -- Discrimination:}
This task explores the user performance when comparing the values given at different time steps. Two time series are shown next to each other, and the user has to select the one with the highest value at the marked point in time. The markers for different time series are at different positions. Again, the time step is indicated with a small marker below the graph. In case of \chg{} and \bhg{} the hue of the marker refers to the hue of the corresponding slice.
Referring to the Triad framework, both the temporal and spatial dimension are given, while the user has to find the relation between the time series
(When + Where $\rightarrow$ What).

\subsubsection*{Task 05 -- Difference estimation:}
The setting of this task is the same as for Task 04.
However, the user has to state the concrete difference in values between two time series at specific time steps. This considers the extended definition of comparison by Andrienko and Andrienko~\cite{andrienko06book}, which encourages to ask for the specific numerical difference between values.

\subsubsection*{Task 06 -- Time estimation:}
A common task of time series analysis is to determine the exact point in time of a certain event. 
In this task we present a small grid of 3 $\times$ 3 time series, where one of the time series is highlighted with a colored bounding box.
We ask the user to specify the point in time of the global maximum in the highlighted graph (What + Where $\rightarrow$ When).

\subsubsection*{Task 07 -- Synoptic search:}
In this task, a grid of 5 $\times$ 5 time series is shown. The user has to select all time series that rise above a certain threshold in a specific time interval (e.g., find all time series that rise above a value of 70 between 0:00 and 8:00). The number of time series that fulfill this criterion lies between 5 and 10. The threshold and the time interval are given in the task explanation of each repetition. Furthermore, the time interval varies within each repetition. The search in this task is synoptic since the user has to consider the entire time interval and compare it with the specified threshold.

\subsubsection*{Task 08 -- Synoptic slope:}
\begin{figure}[htbp]
	\centering
    \includegraphics[width=.9\columnwidth]{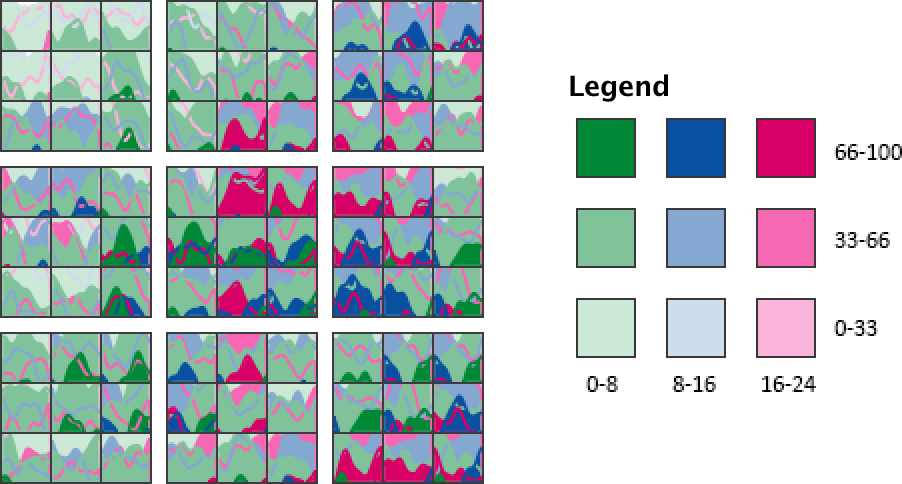}
    \caption{Example of the synoptic slope task from our study: 
``You see 81 images of time series, where groups of $3\times 3$ images form a so-called quadrant. Please click on the quadrant where the time series rise most on average between 8:00 and 16:00.''}\label{fig:synoptic_slope_task}
\end{figure}

This task examines the intermediate reading level in the spatial dimension of the four techniques.
The intermediate level, which is categorized as synoptic according to Andrienko and Andrienko \cite{andrienko06book}, does not ask for an individual element in the set, but rather considers subsets as a whole.
In this task we show 81 time series in a 9 $\times$ 9 grid and define disjoint subsets (so-called quadrants) of 3 $\times$ 3 elements.
The user then has to find the quadrant which has the highest increase on average over a given time interval. 
Hence, the user needs to aggregate the slopes of all time series within a quadrant and compare the results with the other quadrants (see \autoref{fig:synoptic_slope_task}).

\subsubsection*{Task 09 -- Classification:}
In this task, the user is faced with a grid of 5 $\times$ 5 time series, where one time series is highlighted. The user has to state whether the values of the highlighted time series stay within a certain range or exceeds this range compared to the first value of the time series (When + Where $\rightarrow$ What).

\subsubsection*{Task 10 -- Homogeneity:}
The last task again examines the intermediate reading level in the spatial dimension. A grid of 81 elements in a 9 $\times$ 9 grid is shown. A subset of 3 $\times$ 3 elements form a quadrant.
With this setup the user has to compare all quadrants and find the one with the highest homogeneity over the entire time series among its elements (What + When $\rightarrow$ Where).
The homogeneity is determined by the cost of the Dynamic Time Warping algorithm within a quadrant. 

%% file: results.tex
\begin{figure*}[h]%
\centering
\includegraphics[width=\textwidth]{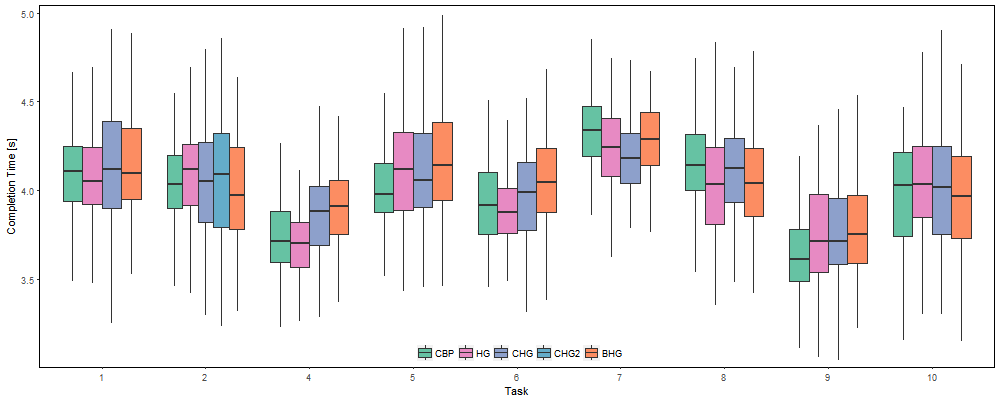}%
\caption{Completion times for all tasks in seconds. Task 3 is integrated in Task 2 and marked as \textit{CHG2}.}%
\label{fig:res_timings}%
\end{figure*}

\begin{figure*}[h]%
\centering
\subfloat[Task 1, 2, 3, 5, 6, 8 and 10]
	{
    \includegraphics[width=0.583\textwidth]{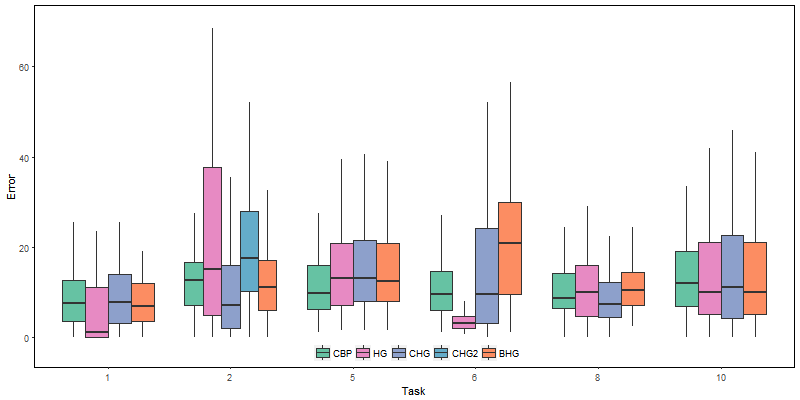} 
	}   ~
	\subfloat[Task 7]
	{
    \includegraphics[width=0.131\textwidth]{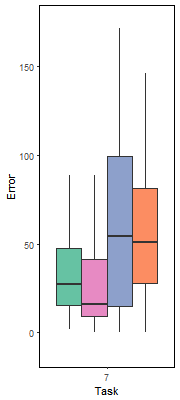} 
	} ~
	\subfloat[Task 4 and 9]
	{
    \includegraphics[width=0.255\textwidth]{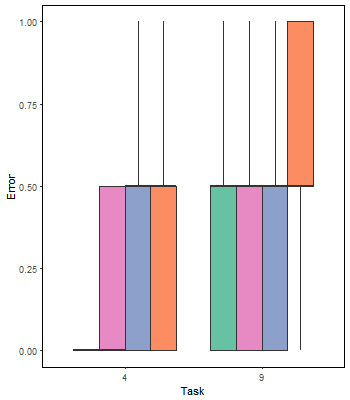} 
	}
\caption{Errors for all tasks. Task 03 is integrated in Task 02 and marked as \textit{CHG2}. Task 07 uses accumulated errors. Task 04 and 09 used binary data, which was averages by the repetitions of each participant.}%
\label{fig:res_errors}%
\end{figure*}

\section{Analysis of results}\label{sec:results}
\begin{figure}[h]%
\centering
\includegraphics[width=\columnwidth]{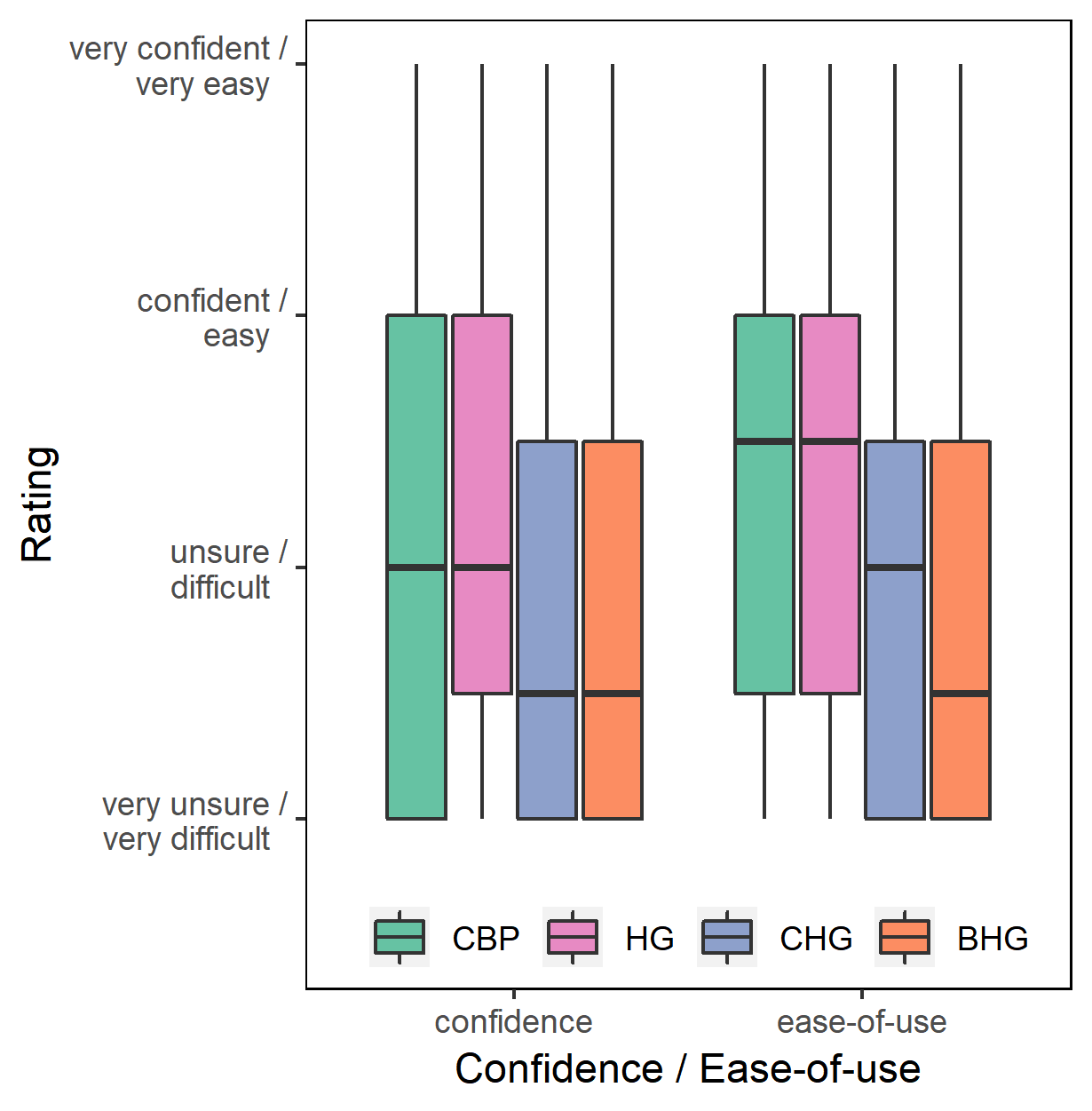}%
\caption{Confidence and ease-of-use of all tasks (more confident / easier on top).}%
\label{fig:res_posttask}%
\end{figure}

In the following, we examine completion time (\autoref{fig:res_timings}), accuracy (\autoref{fig:res_errors}), as well as self-reported confidence and perceived difficulty (both in \autoref{fig:res_posttask}).
In order to facilitate interpretation and generalization of the results, we first report the outcomes of the statistical tests for each task and afterwards discuss the results on a higher level.

\subsection{Analysis approach}
We ran two repetitions per condition (three in case of Task 8) to increase robustness. As recommended by Dragicevic\cite{dragicevic2016stats} we average the results of these repetitions to get one observation per person. 
Histograms and QQ plots were used to visually check the data for normal distribution, whereas Shapiro-Wilk tests were used to confirm this quantitatively. Only the logarithmic transformation of the completion times yields a normal distribution, which enables the use of parametric tests, such as repeated-measures ANOVA and Tukey's HSD test and pairwise t-tests with Bonferroni correction for post-hoc analysis. In the other cases, the non-parametric counterparts were used, Friedman tests and pairwise Wilcoxon tests with Bonferroni correction.

\subsection{Task 01 -- Maximum}
The results of the RM-ANOVA showed no significant effect of the independent variable \textit{technique} on the dependent variable \textit{completion times} ($F(3,291) = 1.61, p = 0.188$).
While all four techniques yield similar completion times,
a Friedman rank sum test revealed that there is a significant difference between the techniques in terms of their accuracy ($\chi^2 = 18.85, p < 0.001$). The following post-hoc analysis showed that \hg{} outperforms \bp{} ($p < 0.05$) and \chg{} ($p < 0.04$). 
The analysis of the reported confidence ($\chi^2 = 45.04, p < 0.0001$) and difficulty ($\chi^2 = 57.2, p < 0.0001$) of this task unveils that participants feel more confident and perceive the task less difficult using either \bp{} or \hg{} compared to \chg{} or \bhg{}.

\subsection{Task 02 -- Increasing slope}
As for the previous task, Task 02 showed no significant differences for the completion times ($F(3,291) = 0.99, p = 0.399$). However, in terms of accuracy there is again a significant difference ($\chi^2 = 16.70, p < 0.001$): Post-hoc analysis reveals that \hg{} performed worse than any other technique (\bp: $p < 0.01$, \bhg: $p < 0.001$, \chg: $p < 0.0001$).
In terms of confidence and task ease, participants reported that for \hg{} (both $p < 0.01$) and \bp{} (both $p < 0.001$) they feel more confident and perceived the task to be easier compared to \chg{} and \bhg.

\subsection{Task 03 -- Decreasing slope}
We compared the results of Task 02 for \chg{} and results of this task. It showed that there is no significant effect regarding the completion times, confidence and task ease. In terms of accuracy, however, there is a strong effect ($p < 0.0001$) when asking for the decreasing slope, which resulted in an average error of $21.28$ compared to $10.97$ for increasing slopes.

\subsection{Task 04 -- Discrimination}
Analysis of Task 04 unveils strong significant effects across all four dependent variables (each $p < 0.0001$). It showed that both \hg{} and \bp{} were faster, more accurate and people felt more confident than with \chg{} and \bhg.

\subsection{Task 05 -- Difference estimation}
With RM-ANOVA, we found a statistically significant effect of \textit{technique} on \textit{completion times} ($F(3,267) = 4.57, p < 0.001$). Pairwise t-tests showed that \bp{} outperformed both \bhg{} ($p < 0.01$) and \hg{} ($p < 0.05$). 
Friedman's test only found a statistical trend of \textit{technique} on \textit{error} ($\chi^2 = 6.43, p = 0.09$).
However with post-hoc tests we found that \bp{} was slightly better than \bhg{} and \chg{} (both $p < 0.05$) 
Neither Friedman nor post-hoc analysis found a statistically significant effect of the participants' confidence.
An analysis of the reported task difficulty showed that there is a difference between the techniques ($\chi^2 = 15.03, p < 0.01$) in general. Pairwise tests revealed differences between \bp{} and \bhg{} ($p = 0.05$) and \hg{} and \bhg{} ($p < 0.05$). In both cases, \bhg{} led to a worse result.

\subsection{Task 06 -- Time estimation}
For the time estimation task, we can report a significant effect of \textit{technique} on \textit{completion times} ($F(3,297) = 7.04, p < 0.001$). \bhg{} required significantly longer completion times than \bp{} and \hg{} (both $p < 0.01$).
Moreover, \hg{} led to results with much higher accuracy than the other techniques (all $p < 0.0001$), followed by \bp{} and \chg. 
This is also aligned with participants' confidence, where \hg{} led to more confident results compared to \chg{} ($p < 0.01$), \bp{} and \bhg{} (both $p < 0.0001$). Besides that, \chg{} performed better than \bhg{} ($p < 0.05$).
In terms of task difficulty, \hg{} also performed significantly better than the other techniques ($p < 0.01$ for \chg{} and \bp{}, $p < 0.0001$ for \bhg{}) and \bhg{} performed worse than the other techniques (\chg: $p < 0.01$, \bp: $p < 0.05$).

\subsection{Task 07 -- Synoptic search}
There was a significant effect of technique on completion time ($F(3,282) = 12.51, p < 0.0001$). Post-hoc analysis showed that \chg{} is significantly faster than \bp{} ($p < 0.0001$) and \bhg{} ($p < 0.001$). Also, \hg{} led to faster completion times compared to \bp{} ($p < 0.01$).
Analysis of the errors with Friedman's test showed that there is a general effect between the techniques ($\chi^2 = 36.02, p < 0.0001$). Pairwise Wilcoxon tests revealed that \hg{} and \bp{} led to better results than \chg{} and \bhg{} (\hg: $p < 0.0001$, \bp: $p <  0.01$).
Analysis of confidence showed that participants using \hg{} felt more confident than with any other technique (\bhg: $p < 0.01$, \bp: $p < 0.01$, \chg: $p < 0.05$).
In terms of reported task difficulty, there was only an effect between \hg{} and \bhg{} ($p < 0.01$) in favor of \hg.

\subsection{Task 08 -- Synoptic slope}
After RM-ANOVA noted a significant effect of technique on completion time ($F(3,294) = 5.99, p < 0.0001$), pairwise t-tests showed that \bp{} is significantly slower than \bhg{} ($p < 0.05$) and \hg{} ($p < 0.0001$).
Comparing the errors showed that \chg{} is more accurate than \bhg{} ($p < 0.05$).
\hg{} led to significantly more confident results compared to \chg{} and \bhg{} (both $p < 0.001$). Also \bp{} resulted in better results than \bhg{} ($p < 0.05$). The same is also valid for task difficulty with \hg{}-\chg{} ($p < 0.05$), \hg-\bhg{} ($p < 0.001$) and \bp-\bhg{} ($p < 0.05$).

\subsection{Task 09 -- Classification}
The RM-ANOVA for Task 09 revealed a statistically significant effect of \textit{technique} on \textit{completion time} ($F(3,297) = 5.66, p < 0.001$). Post-hoc analysis showed that \bp{} led to faster completion times compared to all other techniques (\bhg: $p < 0.001$, \bp, \chg: $p < 0.05$).
Friedman's test also reported an effect of \textit{technique} on \textit{error} ($\chi^2 = 13.85, p < 0.01$). Post-Hoc analysis showed that \bhg{} led to significantly worse results than \bp{} ($p < 0.05$) and \hg{} ($p < 0.01$). 
Analysis of the confidence also revealed a strong differences between the techniques ($\chi^2 = 45.169, p < 0.0001$). Results from pairwise Wilcoxon tests showed that \bp{} led to the highest confidence (\bhg: $p < 0.0001$, \chg: $p < 0.001$, \hg: $p < 0.01$). There is also a significant difference between \hg{} and \bhg{} ($p < 0.001$) with \hg{} having better results.
The same results also apply to task difficulty with \bp{} outperforming every other technique and an additional difference between \hg{} and \bhg.

\subsection{Task 10 -- Homogeneity}
While neither analysis of completion times nor errors revealed any differences, there are significant differences in terms of confidence and difficulty.
It showed that for both cases \bp{} and \hg{} led to better results than \bhg{} or \chg{}.

%% file: discussion.tex
\section{Discussion}\label{sec:discussion}

In this section, we discuss the analysis results and set them in context with our hypotheses.

\textit{Finding \#1: No significant effect of second compression axis on completion times (H1).}
We expected participants to perfom tasks faster using the techniques \hg{} and \bp, which compress only the vertical axis,
while \chg{} and \bhg{} compress both axes.
In our experiment, however, these techniques do not necessarily lead to longer completion times. 
Only Task 04 shows a clear result where both \chg{} and \bhg{} lead to significantly longer completion times.
Task 04 used two different positions, which in case of \chg{} and \bhg{} needed to be mentally assigned to the appropriate time interval, which required additional task time.
Then again, Task 05 uses the same setup as Task 04, but the difference of completion time was weak.
A possible explanation is that naming the relation between two time steps is faster for \hg{} and \bp, but specifying the magnitude of the relation, i.e. the numerical difference between the two time steps, takes longer.
Hence, \textit{H1} can not be confirmed.

\textit{Finding \#2: Similar performance for spatial overview tasks (H2).}
For Tasks 08 and 10, where the question was to find a specific quadrant, we found no conclusive evidence to support \textit{H2}. Given the importance of spatial patterns, we want to investigate this aspect in future work.

\textit{Finding \#3: Data aggregation similarly accurate (H3).}
In all tasks, except Task 01, data aggregation, i.e., \bp{}, did not lead to less accurate results compared to visual compression techniques. Hence \textit{H3} can not be confirmed either.
However, this may be due to the smoothing of the synthetic data generation, which removed high-frequency details.
Also, we only aggregated over a small number of time steps, which should be extended to obtain clearer results.

\textit{Finding \#4: Horizontal collapsing not more accurate for time estimation (H4).}
We expected the techniques \chg{} and \bhg{}, which compress the time axis to yield higher accuracy with longer completion times.
Interestingly, the results of Task 06 rejected \textit{H4} in both dependent variables.
While \bhg{} was found to be significantly slower than \hg{} and \bp{} at time estimation, this was not the case for \chg{}. On the other hand, \autoref{fig:res_errors} reveals that \hg{} is significantly more accurate at time estimation than the other techniques. Deeper investigations would be needed as it was found that the accuracy of \hg{} is much higher than expected.

\textit{Finding \#5: Collapsing allows for more accurate slope estimation (H5).} The results of Task 02 clearly show that the two techniques that apply horizontal collapsing enabled more accurate slope estimations than techniques that shrink the horizontal axis. 
This supports \textit{H5} and goes along with the importance of 
the perception of slopes in a line graph.\cite{cleveland88, Talbot11}

It has also been shown that there is no significant difference between preserved slopes, as it is the case for \chg{} and \bhg, and slopes that are smoothed as in \bp{}.
A possible explanation is that solid areas below a line graph allow for a more accurate slope perception than a single thin line. Another explanation could be a possible drawback of the chosen color maps, since differences in hue are less prominent for the lower bands than higher bands.
The comparison between the results of Task 2 and Task 3 also confirm our hypothesis that the ordering of slices in case of \chg{} emphasizes particular trends in the data (increasing or decreasing slope). Instead of globally specifying the order of slices, one could thus compute the slope of each individual time series and automatically apply the ordering that emphasizes that particular data trend. However, this is subject of future work.

\medskip
The analysis of the results revealed additional outcomes, which were not considered during the formulating the hypotheses.

\textit{Finding \#6: Visual complexity reduces confidence and ease-of-use.}
Throughout almost all tasks (see \autoref{fig:res_posttask}), it has been the case, that \bp{} and \hg{} received higher ratings in the post-task questionnaires while \chg{} and \bhg{} resulted in lower ratings.
Despite considering the novelty of \chg{} and \bhg{}, this leads to the conclusion that a higher visual complexity, i.e. the second axis of compression and possibly also the degree of visual clutter of \chg{} and \bhg{} has a negative effect on the confidence and perceived ease-of-use of the performed tasks.

Analyzing the number of skips per task and technique can give further hints, whether a certain task caused difficulties in general or only for a specific technique. In our case, Task 05 has been skipped at a considerably higher rate among all techniques, but especially in case of \bhg{}, than other tasks. This suggests, that participants had problems, reading the exact difference between two values of different time steps.
In case of Task 09, only \bp{} led to an increased number of skips, which indicates, that it is too difficult for participants to quickly spot threshold-exceeding time series.

\textit{Finding \#7: CBP excels for deviation task.}
Task 09 shows that \bp{} are significantly faster than other techniques, while having a similar accuracy. Task 09 was performed to test the capabilities of the techniques to state the deviation within a time series. The results of \bp{} can be explained in the sense that it is easier to perceive the width of the quartile bands around the median line in combination with the slope of the time series.

%% file: limitation.tex
\section{Limitations and future work}\label{sec:limitations}

Throughout the design process and based on the results of the conducted pilot study, we constantly improved and adjusted the overall design and the specification of tasks. Each task should target a specific question with a precise formulation.
However, the analysis of the results unveiled some design flaws. For example the average accuracy of Task 09 was 44\%, which is almost equal to guessing.
While this result is still valuable in a sense that it shows that none of the four techniques provides the needed accuracy for such a narrow range, the goal of the task was to find a more precise answer, which speaks in favor for a specific technique.
Increasing the allowed deviation from the first value of the time series in Task 09 would possibly lead to a more meaningful result. The currently chosen value seems to be close to the noticeable limit.

During the design of \chg{} and \bhg{}, the advantages and disadvantages of different bivariate color maps have been extensively discussed. The current sequential--qualitative color map provides a good distinction between the value and the temporal axis while still emphasizing the temporal progression. However, the difference in hue between the lower bands is relatively small, which makes them difficult to distinguish.

Further, negative values were discussed during the design, but were discarded due to the increased number of necessary colors introduced by a diverging color map: Each slice would require an additional color hue to display negative values, which in our case would result in total 6 different hues. Additionally, we did not find corresponding positive-negative color pairs, that could facilitate the examination of the graph.
Thus, this topic was out of the scope of this study, but opens interesting design challenges for the future.

\autoref{tab:task-cat1} shows that our set of tasks covers nearly all areas of the spatial and temporal dimensions. However, most of the tasks focus on the temporal aspect of the data. Therefore it would be interesting to explore the performance of the techniques in terms of spatial pattern detection. Moreover, a gradual increase of the spatial domain beyond quadrants would yield further insights of the applicability of aggregation and color encoding for time series visualization. It would be interesting to find transition points at which the performance of these techniques significantly increases or decreases and thus other approaches need to be considered.

It has been shown that the visual complexity of \chg{} and \bhg{} lowers the confidence of the participants' answers and reduces the ease-of-use of the tasks. A follow-up question would investigate how additional training and domain knowledge can affect these results.

We also want to investigating the effect of user interaction with the techniques such as adjusting the number of aggregated time steps for \bp{} with high frequent data and specifying the order of slices of \chg{} to emphasize a different time interval (see \autoref{fig:rl_lr}). Moreover, the original line graph could be shown on mouse-over as details-on-demand.

%% file: conclusion.tex
\section{Conclusion}

In this article we investigated the advantages and disadvantages of data aggregation and color encoding of time series data within a dense spatial context. Further, we examined whether the advantages of the horizontal collapsing of \hg{} can also be exploited in the temporal axis and prove to be beneficial. This question led to the experimental designs of \chg{} and \bhg. To assess the performances of four different techniques we conducted a quantitative evaluation.
To systematically cover different spatio-temporal aspects, we have created a set of tasks based of two conceptual frameworks \cite{andrienko06book, peuquet94triad}.
The results of the user study show that the different techniques entail different strengths and weaknesses: 
\begin{itemize}
\item \hg{} are most accurate in comparing extrema among multiple time series and specifying the time of a certain event, but are inferior in slope tasks due to the effects of distortion. 
\item \bp{} do not excel in a certain task or group of tasks, but provide similar accuracy across all tasks. 
\item our proposed extensions \chg{} and \bhg{} have shown to be comparably effective in certain aspects to the other techniques, while preserving the horizontal resolution and details of the original line graph. Therefore, they provide better accuracy for slope tasks, but show contradictory result in lower accuracy when specifying the point in time of a certain event. 
\item \chg{} emphasize the temporal progression of the time series, but also suffer from occlusion, especially in static application. 
\item \bhg{} avoid this occlusion with the concept of braiding, but increase cognitive load and visual complexity.
\end{itemize}